\documentclass[aip,prl,preprint]{revtex4}
\usepackage{graphicx}
\bibstyle{nature}	
\usepackage{color}

\begin{document}

\title{Direct band structure measurement of a buried two-dimensional electron gas}

%

\author{Jill A. Miwa}
\affiliation{Department of Physics and Astronomy, Interdisciplinary Nanoscience Center (iNano), Aarhus University, 8000 Aarhus C, Denmark.}
\author{Philip Hofmann}
\affiliation{Department of Physics and Astronomy, Interdisciplinary Nanoscience Center (iNano), Aarhus University, 8000 Aarhus C, Denmark.}
\author{Michelle Y. Simmons}
\affiliation{Centre of Excellence for Quantum Computation and Communication Technology, School of Physics, University of New South Wales, Sydney, NSW 2052, Australia. }
\author{Justin~W.~Wells}\email[]{quantum.wells@gmail.com}
\affiliation{MAX IV Laboratory, Lund University, Sweden.}
\affiliation{Department of Physics, Norwegian University of Science and Technology (NTNU), N-7491 Trondheim, Norway }
\date{\today}


\maketitle


\textbf{Buried two dimensional electron gasses (2DEGs) have recently attracted considerable attention as a testing ground for both fundamental physics and quantum computation applications \cite{Fuechsle:2012,Weber:2012,Mahapatra:2011}. Such 2DEGs can be created by phosphorus delta ($\delta$) doping of silicon, a technique in which a dense and narrow dopant profile is buried beneath the Si surface. Phosphorous $\delta$-doping is a particularly attractive platform for fabricating scalable spin quantum bit architectures \cite{Fuechsle:2012,Weber:2012,Mahapatra:2011}, compatible with current semiconductor technology. The band structure of the $\delta$-layers that underpin these devices has been studied intensely using different theoretical methods \cite{Carter:2009a,Carter:2011,Drumm:2012,Lee:2011b,Ryu:2009,Qian:2005}, but it has hitherto not been possible to directly compare these predictions with experimental data. Here we report the first measurement of the electronic band structure of a $\delta$-doped layer below the Si(001) surface by angle resolved photoemission spectroscopy (ARPES). Our measurements confirm the layer to be metallic and give direct access to the Fermi level position. Surprisingly, the direct observation of the states is possible despite them being buried far below the surface. Using this experimental approach, buried states in a wide range of other material systems, including metallic oxide interfaces \cite{Ohtomo:2004,Jang:2011,Yoshimatsu:2011}, could become accessible to direct spectroscopic investigations. }\\

\begin{figure}
\includegraphics[width=0.9\textwidth]{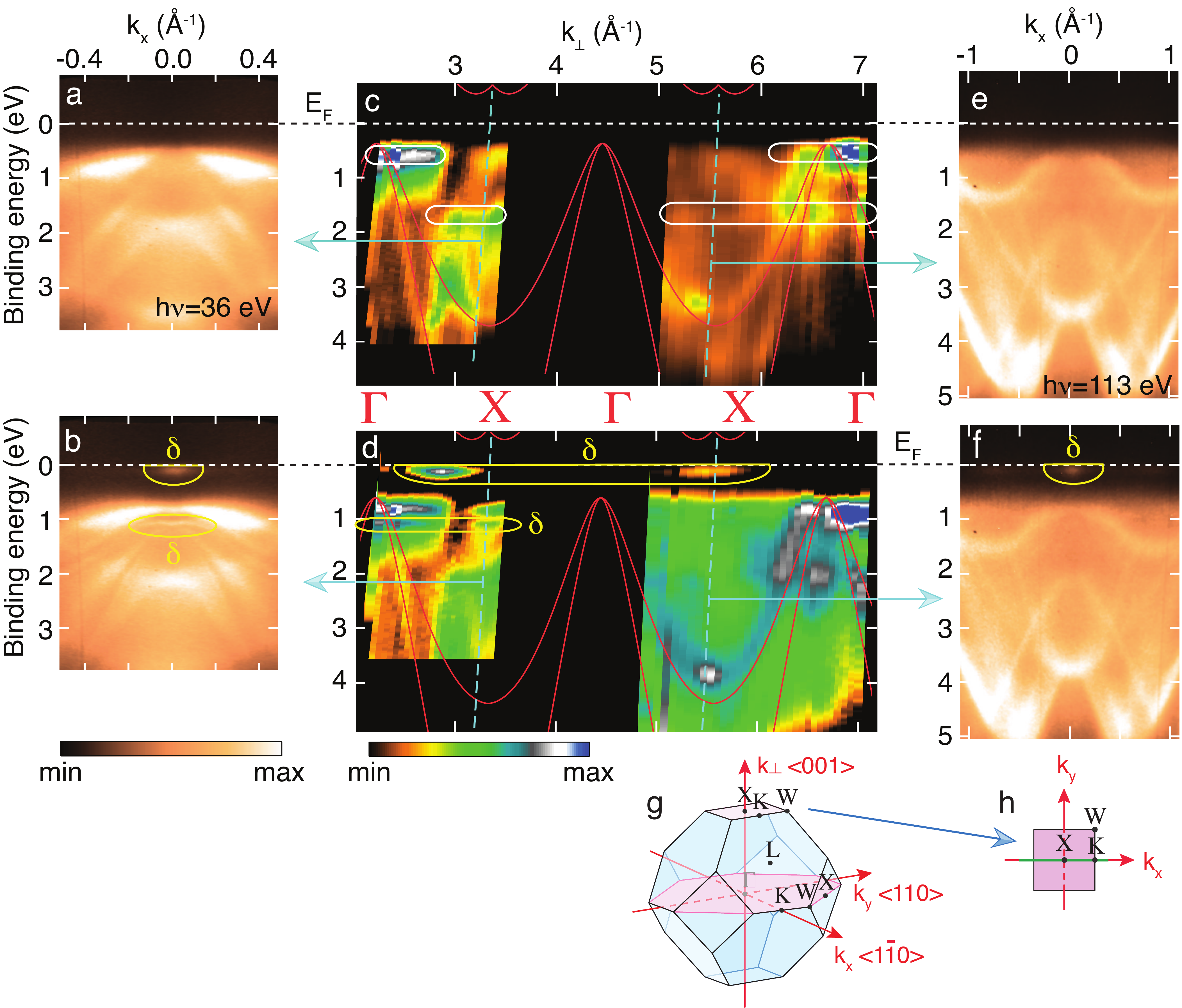}
\caption{\textbf{Band structure measurements of buried phosphorus $\delta$-layers in Si(001).} \textbf{a} and \textbf{b} ARPES measurements at $h\nu=36$~eV for the control sample and $\delta$-doped silicon respectively. The additional states due to the $\delta$-layer are enclosed in yellow and marked. \textbf{c} and \textbf{d}; the photoemission intensity at normal emission, plotted as a function of $k_\perp$ for the same samples with the calculated bulk band structure \cite{Rohlfing:1993} overlaid in red. $\delta$-layer states are marked as before and surface states are enclosed in white in \textbf{c}. ARPES measurements taken at $h\nu=113$~eV for the control sample \textbf{e} and the $\delta$-doped sample \textbf{f}. The bulk Brillouin zone \textbf{g} and the definition of the axes $k_\perp$, $k_x$ and $k_y$, with the plane through the bulk $X$ point extracted \textbf{h}. The green solid line along $k_x$ corresponds to the bulk directions seen in the ARPES measurements in \textbf{a, b, e} and \textbf{f}.
\label{megafigure}}
\end{figure}

In order to probe the electronic structure of the $\delta$-layer states, samples were prepared by substitutionally incorporating 1/4 of a monolayer (ML) of phosphorus atoms on a clean  Si(001) surface and burying this structure under 2.0$\pm0.6$~nm of silicon \cite{suppl_mat}. Control samples were fabricated by similarly growing silicon on the clean surface, without the incorporation of a phosphorus layer. Fig. \ref{megafigure} a,e and b,f show ARPES measurements at two photon energies (36 and 113~eV) of a control sample and a $\delta$-doped sample, respectively. Both samples show several similar  structures that can be ascribed to silicon bulk and surface states, but the $\delta$-doped sample clearly shows sharp additional states that are missing on the control sample (enclosed in yellow).  Most strikingly, we observe a state at the Fermi energy ($E_b<0.2$~eV), immediately confirming the predicted metallic nature of the $\delta$-doped layers. Other additional states are found at a binding energy $ E_b \approx$ 1~eV with a character similar to the bulk $\Gamma$ valence band (VB) states.

ARPES can probe the dimensionality of the observed states by acquiring data at different photon energies  \cite{suppl_mat}. Varying the photon energy corresponds to a change in the wave vector perpendicular to the surface ($k_\perp$) and while bulk states generally show dispersion as a function of $k_\perp$, the two-dimensional states in the $\delta$-layer or genuine surface states do not. Fig. \ref{megafigure} c and d show the photoemission intensity in normal emission ($k_x=k_y=0$) for the control sample and $\delta$-doped sample, respectively,  across the available photon energy range (14-200~eV). The horizontal axis has been converted from photon energy to  $k_\perp$ assuming free electron final states \cite{suppl_mat}.  The marked states in  Fig. \ref{megafigure} b and f are clearly identified as two-dimensional by their lack of dispersion. Furthermore, they can be distinguished from the genuine surface states or surface resonances (emphasised by white contours at $E_b\approx$0.5 and 2.0~eV) of Si(001), since these are present on both samples. Note that the surface states appear quite broad. This is ascribed to crystalline imperfection in the surface of the overlayer, grown at low temperature to minimise dopant segregation.

The ARPES measurements in Fig. \ref{megafigure} c and d are plotted together with the calculated many-body bulk band structure for Si  (after Ref. \cite{Rohlfing:1993}) and the agreement is  very good. The $k_\perp$ values for reaching the bulk $X$ high symmetry point are found to correspond to photon energies of 36 and 113~eV. Although the bulk states are essentially unchanged by $\delta$-doping, the introduction of a layer of $n$-type dopants has shifted the surface Fermi level toward the top of the band gap by $\approx0.23$~eV. This results in a shift of the entire band structure (shallow bulk and surface states) to a higher binding energy as seen, for example, by comparing Fig.\ref{megafigure} a and b. 

\begin{figure*}
\includegraphics[width=\textwidth]{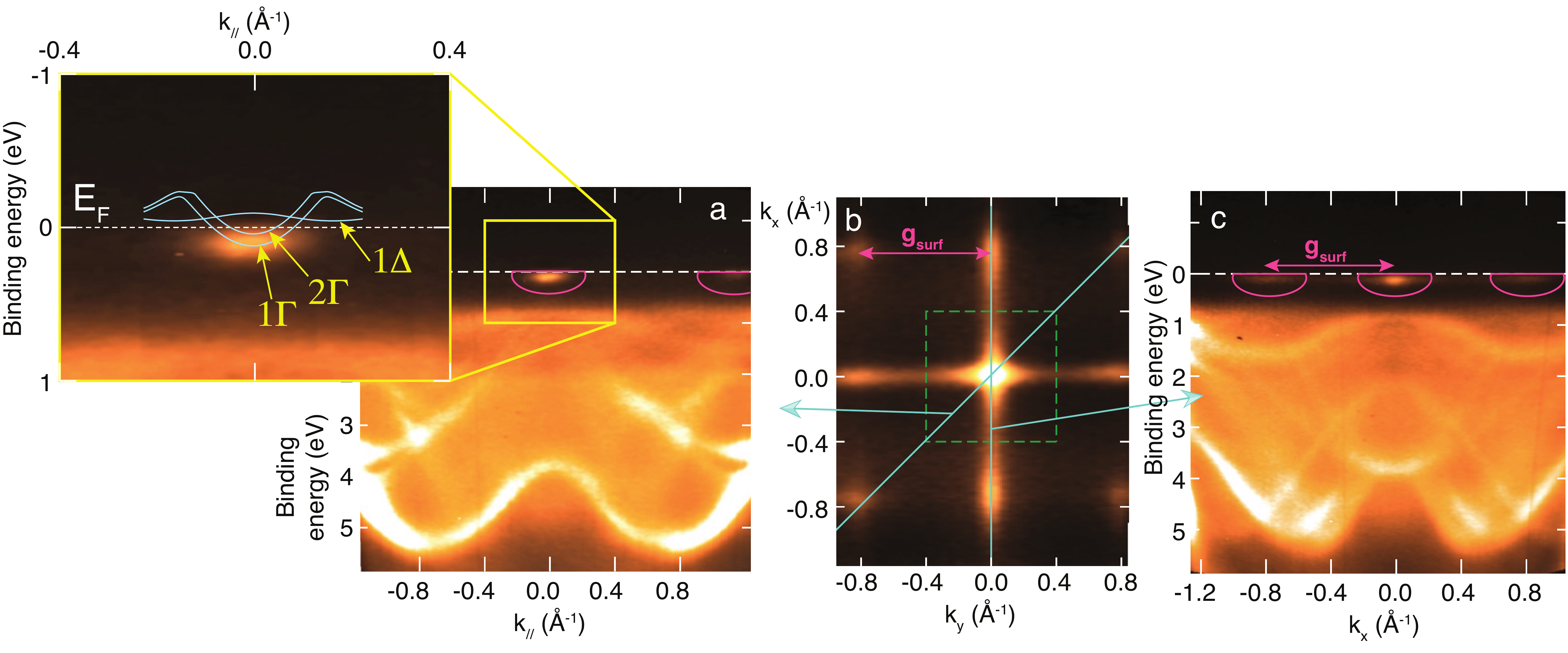}
\caption{\textbf{Periodicity of $\delta$-layer states.} \textbf{a} and \textbf{c} ARPES measurements along $k_x=k_y$ and $k_y=0$, respectively, for the $\delta$-doped sample at $h\nu=113$~eV with the $\delta$-layer states enclosed in pink. The inset shows the detail of the $\delta$-layer state with the calculated $1\Gamma$, $2\Gamma$ and $1\Delta$ states (after Fig. 10 d of Ref. \cite{Carter:2011}) overlaid. \textbf{b} A constant energy map at the Fermi level. The $2\times2$ surface unit cell is indicated by a green dashed box and the pale blue lines and arrows indicate where the band maps \textbf{a} and \textbf{c} have been extracted. The length of the surface reciprocal lattice vector $\mathbf{g_{surf}}$ is indicated.} 
\label{fermi}
\end{figure*}

 
The new electronic states in the $\delta$-doped sample can be mapped in the plane parallel to the surface. Fig. \ref{fermi} shows the dispersion in two high-symmetry directions (a and c), as well as the surface Fermi contour (b). The map directly reveals the periodicity of the dispersion parallel to the surface as corresponding to a $(2 \times 2$) structure. This is to be expected because of the two rotational domains of $(2 \times 1)$ reconstruction present on a Si(001) surface. Note that this periodicity parallel to the surface does not necessarily imply the same periodicity in the $\delta$-layer or in the bulk, it is anyway expected due to the well-known process of surface-umklapp scattering   \cite{suppl_mat}.  The Fermi surface contour appears to be streaked along the $k_x$ and $k_y$ directions. The streaks can be related to the different appearance of the state along the two cuts shown. In the $k_x=k_y$ direction in  Fig. \ref{fermi} a, the $\delta$-layer state appears as a sharp structure around $k_{\parallel}= (k_x^2+k_y^2)^{0.5} = 0 $, whereas it is always observable along the $k_x=0$ and $k_y=0$ (Fig. \ref{fermi} c) directions. The detailed reason for this is difficult to determine because the states are only barely visible at the Fermi energy and the detailed dispersion is not accessible. The dispersion could be anisotropic with a higher effective mass along $k_x$ and $k_y$ than along the diagonal direction. However, streaking would also be expected if step edges or the rotated $2 \times 1$ domains on the surface are restricted in the $x$ and $y$ directions. Similar limited domain sizes in the $\delta$-layer could also play a role but the present observations do not require the existence of new periodicities (such as $4\times n$) in the delta layer \cite{Carter:2009a}.

Theoretical studies of $\delta$-doped Si have primarily focused on the electronic states close to the Fermi energy, as these are the relevant states for transport \cite{Carter:2009a,Carter:2011,Drumm:2012,Lee:2011b,Ryu:2009,Qian:2005}. A common prediction of the studies is the formation of  impurity bands below the bulk conduction band minimum. The lowest lying states, called $1\Gamma$ and $2\Gamma$, derive from the projection of the conduction band minimum states onto the plane of the $\delta$-layer. These states are non-degenerate, and different theoretical approaches have resulted in different calculated separation energies between them at  $k_{\parallel}=0$, referred to as valley splitting (see, for example, Refs. \cite{Carter:2009a,Carter:2011}). Even for the present case of 1/4~ML, the Fermi level position of the $1\Gamma$ band minimum and the valley splitting are dependent on factors such as the dopant distribution \cite{Carter:2011}.  These values also vary between theoretical approaches, however for the present case of 1/4~ML binding energies of 300-500~meV for the $1\Gamma$ band minima are typical \cite{Carter:2011}.

Our observed metallic states can be assigned  to the $1\Gamma$ states and possibly also the $2\Gamma$ states, depending on the size of the valley splitting. The observed occupation of the states is, however, much lower than predicted and appears to be $<200$~meV. The most important reasons for this discrepancy is that the calculations assume an intrinsic Si bulk, in contrast to our strongly $p$-doped samples, and a symmetric potential around the $\delta$-layer. It is nonetheless interesting to overlay calculated $\delta$-layer states on the experimental data (inset of Fig. \ref{fermi} a) -- for the purpose of illustration, the $1\Gamma$, $2\Gamma$ and $1\Delta$ states adapted from Fig. 10 d of Ref. \cite{Carter:2011} are assumed. The Fermi level has been shifted by 140~meV in order to facilitate the comparison, but other than this the axes are the same. The width, depth and general shape of the measured state is consistent with the calculation. However, only one state is observed. This could be due to the limited experimental resolution or to a valley splitting that is slightly larger than calculated such that the $2\Gamma$ state remains occupied. In any event, it appears safe to assume that the next band, the $1\Delta$ state, is sufficiently separated in energy to be unoccupied and hence to preclude its observation.

\begin{figure*}
\includegraphics[width=\textwidth]{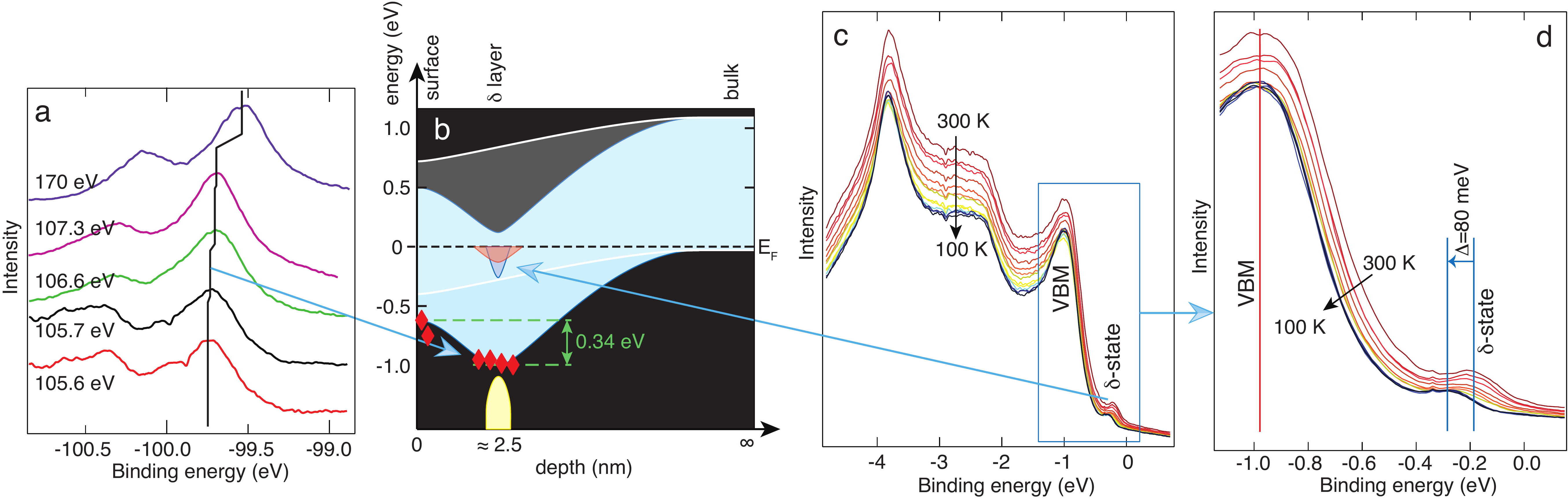}
\caption{\textbf{Band alignment and temperature dependence of the $\delta$-layer states.}  \textbf{a} Low kinetic energy measurements of the Si~2p core level (after subtraction of a polynomial background) at normal emission for a range of photon energies. $h\nu=105.6$~eV corresponds to a $E_k$ of just $\approx$1.3~eV. \textbf{b} Schematic representation of the band bending; the pale blue band indicates the silicon band gap.  The red diamonds show the VBM position as extracted from the core level measurements (uncertainty in energy $<50$~meV). The $\delta$-layer state within the band gap is indicated in red ($T$=300~K) and blue (100~K). The $\delta$-layer state at deeper binding energy is indicated schematically in yellow. The band bending for the same substrate, without the $\delta$-layer is overlaid as a transparent white band. \textbf{c} Temperature dependent measurements of the valence band at normal emission and $h\nu=113$~eV, and \textbf{d} the same data expanded to show the detail of the $\delta$-layer state near the Fermi level and VBM.  }
\label{bandbending_t_dep}
\end{figure*}

The $\delta$-layer states at a binding energy of $\approx$1~eV (Fig. \ref{megafigure} b) are spectroscopically quite similar to quantum-confined hole states  \cite{Takeda:2005} but they are interpreted as valence band-derived impurity bands. A comparison to theoretical work is not possible at present for lack of published calculations in this energy range. These states might, however, give valuable information on the electronic structure of the $\delta$-layers, even though they do not contribute to the actual transport. 


It is surprising that the electronic structure of the delta layers should be observable by ARPES at all.  A simplistic view of photoemission suggests that the mean free path of the photoelectron is so short ($<$1~nm) that the photoemission intensity becomes vanishingly small for layers buried by just a couple of unit cells \cite{Monkman:2012}. Indeed, this is what happens to the photoemission intensity from the P~2p core level that is  entirely attenuated after 6 minutes of Si dosing, indicating negligible segregation of the dopants  \cite{Goh:2004,Goh:2009} and allowing us to calibrate the Si overlayer thickness to be at least 1.4 nm for this evaporation time. The $\delta$-layer states in the valence band and conduction band, on the other hand, are still observable for Si overlayer thicknesses in excess of 5~nm. The reason for this behaviour is the extended nature of these $\delta$-layer states. While they are quite strongly confined to the region close to the $\delta$-layer \cite{Lee:2011b}, their wave function still has a very small amplitude at the surface. This is sufficient to observe the states when they are resonantly enhanced by choosing the right photon energy. Such resonances are clearly observed in  Fig. \ref{megafigure} d, where the states at the Fermi energy are most intense for photon energies that lead to emission near the bulk $X$ points. This is consistent with the fact that the $1\Gamma$ and $2\Gamma$ states are derived from the conduction band minimum. Such strong resonances in cross section are frequently observed for surface states \cite{Louie:1980} and surface 2DEGs \cite{Colakerol:2006,King:2010,Bianchi:2010b} and permit the observation of states that have a wave function maximum far below the surface.

Being able to directly probe the states of the $\delta$-layer, relative to the surface and bulk states allows the band alignment in this region to be measured. As mentioned above, the position of the Fermi level at the surface is directly measured and the presence of the $\delta$-layer leads to a shift of $\approx0.23$~eV with respect to the surface of uniformly doped Si \cite{Maartensson:1986}.  The position of the $\delta$-layer states themselves where found to be independent from the thickness of the Si overlayer for a thickness range of nominally $2$ to $5$~nm (not shown). Note, however, that there is a significant uncertainty in the absolute Si overlayer thickness \cite{suppl_mat}. The band alignment on the bulk side of the $\delta$-layer cannot be directly probed here for these thicknesses, but it is reasonable to expect alignment back to the bulk Fermi level (which is straightforward to calculate \cite{Himpsel:1983a}). A schematic representation of the room temperature band alignment is given in Fig. \ref{bandbending_t_dep} b. Note that this scenario is more complex than the situation usually assumed in calculations of the $\delta$-layer's properties \cite{Carter:2009a,Carter:2011,Drumm:2012,Lee:2011b,Ryu:2009,Qian:2005}. First of all, the bulk Si used is strongly $p$-doped, fixing the Fermi level on one side of the potential well. The Fermi level at the surface is largely dictated by the surface states (even though the presence of the $\delta$-layer leads to a shift), such that the near surface region is seen to be $n$-type. Such details are important for the theoretical modelling of the $\delta$-layer's electronic structure and ultimately for the operation of $\delta$-layer based devices \cite{Fuechsle:2012,Mahapatra:2011,Weber:2012}. 


The schematic band bending near the surface can be confirmed by core level spectroscopy. To this end, Si~2p core level spectra were measured as a function of photon energy and thus electron kinetic energy $E_k$ (Fig. \ref{bandbending_t_dep} a). Since the mean free path $\lambda$ of the photo-emitted electron has a dependence on $E_k$ \cite{Seah:1979}, the sampling depth is also dependent on $E_k$, with $E_k\approx70$~eV corresponding to a minimum in $\lambda$ of $\approx0.6$~nm. A large increase in $\lambda$ can be achieved with very low $E_k$ \cite{Himpsel:1983a}.  It is possible to collect the Si~2p core level from $E_k<2$eV (sampling depth $\approx$ 2.5~nm \cite{Himpsel:1983a}) to $E_k>70$eV (sample depth $\approx$ 0.6~nm), corresponding to depths within the region of strongest band bending, as illustrated in Fig. \ref{bandbending_t_dep} b. Although there is significant uncertainty in estimating the sampling depth, this allows measured values of the band bending to be overlaid on Fig. \ref{bandbending_t_dep} b (red diamonds). 


As an example of the possibilities opened by the direct measurement of $\delta$-layer states, we report the temperature dependence of the state close to the Fermi level. Fig. \ref{bandbending_t_dep} c and d reveal that this state moves to higher binding energy as the temperature is reduced. Between 100 and 300~K, the $\delta$-layer state is seen to shift (reversibly) by around 80~meV, whilst the valence band maximum is essentially unchanged. This observed behaviour is somewhat different from the prediction of an almost constant position with respect to the Fermi level \cite{Ryu:2009}, something that is probably caused by the aforementioned differences in bulk doping and shape of potential well between our experiment and most calculations. The important point is that the experimental approach presented here allows this temperature dependence to be directly measured, opening the possibility to test different calculational approaches.

We emphasise that the experimental approach presented here is not limited to buried phosphorus dopants in silicon.  $\delta$-doping of other semiconductors (such as GaAs \cite{Nazmul:2005} and Ge \cite{Scappucci:2012a}), boron ($p-$type) $\delta$-layers \cite{Mattey:1992} as well as metal oxide interfaces \cite{Ohtomo:2004,Jang:2011,Monkman:2012} could be similarly studied. In addition to a plethora of such materials, a wealth of further information should also be extractable from such experimental data; we have already been able to discuss the temperature dependence to the $\delta$-layer pinning, but more detailed studies should reveal parameters such as the effective mass \cite{Drumm:2012} and the strength of the electron-phonon interaction \cite{Hofmann:2009b,Gayone:2005}.

\section*{Acknowledgements}  
We thank Craig Polley, Oliver Warschkow, Nigel A. Marks, Damien J. Carter and Thiagarajan Balasubramanian for valuable comments and discussions and Johan Adell, Stefan Wiklund and Kurt Hansen for technical support. J.A.M acknowledges financial support from the MAX IV laboratory and the Lundbeck Foundation via Liv Hornek\ae r. This research was conducted in collaboration with the Australian Research Council Centre of Excellence for Quantum Computation and Communication Technology (project number CE110001027) and the US National Security Agency and the US Army Research Office under contract number W911NF-08-1-0527. M.Y.S. acknowledges an ARC Federation Fellowship. 


\newpage

\section*{Supplementary Information}
\subsection*{Experimental details}

ARPES measurements were performed at beamline I4 at MAXlab, using a Specs Phoibos electron analyser equipped with an imaging detector.  The detector's angular imaging direction corresponds to $k_x$. Mapping orthogonal to this ($k_y$) is achieved by rotating the sample. The beamline has photon energy range 14 to 200~eV, but the extremely high cross-section of the Si~2p core levels ($E_b\approx100$~eV) results in core-level excitation from second order light hindering measurements at photon energies of $\approx$100~eV. The beamline and analyser were set such that the energy and angle resolutions were $\approx$50~meV and $\approx$0.3$^{\circ}$, respectively. The binding energies quoted are relative to a metal Fermi level. All measurements were carried out at room temperature, except those in the temperature scan of Fig. 3. For measuring the Si~2p core level at very low kinetic energy, as reported in Fig. 3 a, it was necessary to apply a bias voltage to the sample. Comparing the intensity of data taken at different photon energies (Fig. 1 c and d) necessitates a normalization of the spectra. To present the data in Fig. 1, each normal emission spectrum was normalized to the the total intensity of the detector image. 

Two types of samples are prepared, the $\delta$-doped samples and the control samples. The starting point for both is a  $p$-type Si(001) (0.01 $\Omega$~cm) wafer. This  is prepared \textit{in vacuo} by degassing at 650$^{\circ}$C followed by annealing to 1200$^{\circ}$C, producing a clean and well-ordered surface with two rotational $2\times1$ domains. In order to prepare the control sample, an overlayer ($\approx$1.4 to 5~nm) of Si is then evaporated at a rate of 0.33~nm/min onto the surface held at 250$^{\circ}$C. The deposited overlayer is not sufficiently well-ordered for the band structure to be observed with ARPES. However, a subsequent anneal to $\approx$450$^{\circ}$C reorders the overlayer such that bulk-like bands become visible. The surface states appear suppressed and broad compared with the initial surface (prior to overlayer growth) but the surface and bulk electronic structure of the samples prepared in this way is otherwise similar to those of the initial Si(001)--($2\times1$) surface. 

The $\delta$-doped samples are prepared by depositing phosphorus on the clean Si(001) surface prior to the growth of a Si layer. To this end, phosphine gas, PH$_3$, is dosed (1.4~L) with the substrate at room temperature. The sample is then annealed to 350$^{\circ}$C for 60~s to substitutionally incorporate phosphorus, P, into the surface layer. This procedure leads to a coverage of 1/4 monolayer of phosphorus atoms [S1,S2]. 
A Si layer is then grown on top and annealed as before, creating the buried $\delta$-layer. 

The thickness of the final Si layer is estimated by monitoring the intensity of the P~2p core level, measured at a photon energy of 170~eV, as a function of Si dose. The complete suppression of intensity from this level is estimated to correspond to a minimum layer thickness of 1.4~nm. While the absolute thickness is error-prone, the relative thickness of the preparation is well controlled, and samples with overlayers of 150$\%$  and 250$\%$ of this minimum thickness were prepared. The $\delta$-layer states are observed for all of these samples and their position is unchanged.

\subsection*{Probing bulk states using ARPES}

For the reader not familiar with ARPES, we briefly summarise some of the main relevant points for photoemission from bulk states. The photoemission process for bulk electrons changes the perpendicular component of the electron's wave vector, $k_\perp$, upon crossing the surface barrier. Indeed, strictly spoken $k_\perp$ is not even well-defined near the surface. Without the knowledge of the final state dispersion in the solid, the initial $k_\perp$ of the photoelectrons cannot be determined.

 A frequently used approach to solve this problem is the assumption of free electron final states {[S3]}. 
 To consider a simplified case, we can assume that the effective mass of the final state band in the crystal is equal to the electron mass ($m$), and consider only photoemission along the surface normal (i.e. $k_x=k_y=0$). In this case, it is straightforward to show that $k_{\perp}\approx \sqrt{(2m/\hbar^2)[V_0+h\nu-E_b-\Phi]}$, where $V_0$ is the inner potential, $h\nu$ is the photon energy, $E_b$ is the binding energy of the electron and $\Phi$ is the workfunction. All of these parameters are known [S4] 
 and this approach has been used to convert the horizontal axis in Fig. 1 c and d from photon energy to $k_\perp$. This conservation also identifies the photon energies of 36 and 113~eV to correspond to emission from the bulk $X$ states at normal emission. 
 
 \begin{figure}
\includegraphics[width=\textwidth]{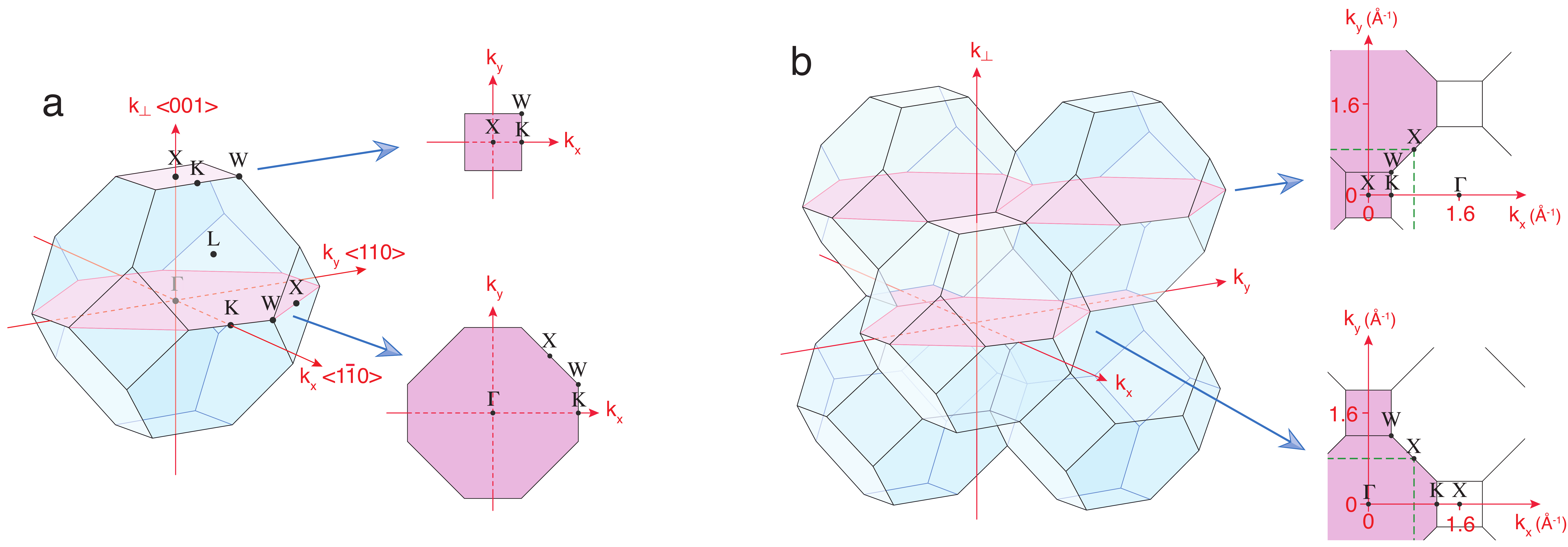}
{\begin{flushleft}
{Supplementary Figure 1: \textbf{Bulk Brillouin zones of Si(001).} \textbf{a} Silicon Brillouin Zone (BZ) \textit{defining} the axes $k_x$, $k_y$ and $k_\perp$. The bulk Brillouin zone is shown, with the axes definitions corresponding to the principle vectors of the primitive surface unit cell. Two planes with fixed $k_\perp$ corresponding to the BZ zone boundary and centre respectively are also shown. \textbf{b} several of the neighbouring BZs are also included such that the continuation of these plans into the neighbouring zones can be seen. The surface ($2\times2$) reciprocal unit cell is overlaid in green (dashed square), such that the implications of momentum exchange due to surface umklapp processes are apparent. }
\end{flushleft}}
\label{si_BZ}
\end{figure}

 The wave vector parallel to the surface $k_{\parallel}$, on the other hand, is conserved in the photoemission process such that $k_{\parallel}$ in the panels of Fig. 1 a, b, e, f and Fig. 2 a and c actually correspond to the directions indicated in the figures. For surface states, surface resonances and two-dimensional $\delta$-layer states, $k_{\parallel}$ is the only relevant quantum number but for the photoemission from bulk states it is important to note that $k_\perp$ changes along the scan. Consider Supplementary Fig. 1a: For a photon energy  of 36~eV or 113~eV, the plane probed by ARPES is curved, with the bulk X-point corresponding to normal emission. The bulk direction probed in Fig. 1 a, b, e, f and 2 c can be approximated to the bulk $X-K$ direction and Fig. 2 a to the $X-W$ direction. It is important to notice that when we refer to $\Gamma$, $X$, $K$ and $W$, we refer to the high symmetry points in the bulk Brillouin Zone, as indicated in  Supplementary Fig. 1 -- this should not be confused with the 2-dimensional Brillouin Zone indices, for example in Ref. [S5]. 
 
Another important point is that, due to the two-dimensional periodicity, $k_{\parallel}$ is only defined plus or minus a surface reciprocal lattice vector $\mathbf{g_{surf}}$ (and we choose to \textit{define} the $x$ and $y$ axes to be in the $1\bar{1}0$ and $110$ directions, such that $\mathbf{g_{surf}}$ is in the directions $\pm\mathbf{\hat{x}}$ and $\pm\mathbf{\hat{y}}$). To see the consequences of this for the photoemission of bulk states from a Si(001) surface, consider Supplementary Fig. 1b which shows an extended Brillouin zone scheme. Variable photon-energy photoemission at normal emission probes the $\Gamma-X$ direction of the bulk band structure and using the free electron final state model, the positions of the high symmetry points are indicated in Fig. 1 c and d. When the bulk $X$ point is reached, however, adding a surface reciprocal lattice vector (within the red plane) folds in bulk states from the $\Gamma$  point of the neighbouring zone. This so-called surface umklapp process is the reason why bulk $\Gamma$ states are frequently observed at photon energies corresponding to the bulk $X$ points and vice versa [S6]. 
This is also evident in the present data, e.g. in Fig. 2 c for the bulk $X$ point reached at $k_\perp \approx 3.5$~\AA$^{-1}$ where a replica of the bulk valence band maximum is clearly observed.

Finally, we note that the surface umklapp processes in this particular case are not limited to the $(1 \times 1)$ reciprocal lattice of Si(001) but to the two rotated domains of a $(2 \times 1)$ reconstruction. This alone would give rise to the observed $(2 \times 2)$ periodicity of the $\delta$-layer states in Fig. 2 b. This observed periodicity does therefore not necessarily reflect the periodicity in the phosphorus $\delta$-layer. \\


%
\noindent 
[S1] H.~F. Wilson, O.~Warschkow, N.~A. Marks, N.~J. Curson, S.~R. Schofield,
  T.~C.~G. Reusch, M.~W. Radny, P.~V. Smith, D.~R. McKenzie, and M.~Y. Simmons.
\newblock Thermal dissociation and desorption of $\mathrm{P}{\mathrm{H}}_{3}$
  on Si(001): A reinterpretation of spectroscopic data.
\newblock {\em Phys. Rev. B}, 74:195310, 2006.\\

\noindent 
[S2] S.~R. Schofield, N.~J. Curson, M.~Y. Simmons, F.~J. Ruess, T.~Hallam,
  L.~Oberbeck, and R.~G. Clark.
\newblock Atomically precise placement of single dopants in Si.
\newblock {\em Phys. Rev. Lett.}, 91(13), 2003.\\

\noindent 
[S3] F.~J. Himpsel.
\newblock Experimental determination of bulk energy band dispersions.
\newblock {\em Appl. Opt.}, 19(23):3964--3970, 1980.\\

\noindent 
[S4] D.~E. Ashenford and N.~D. Lisgarten.
\newblock The measurement of inner potential for diamond, germanium and
  silicon.
\newblock {\em Acta Crystallographica Section A}, 39(3):311--314, 1983.\\

\noindent 
[S5] D.~J.~Carter, O.~Warschkow, N.~A.~Marks, and D.~R.~McKenzie.
\newblock Electronic structure models of phosphorus $\delta$-doped silicon.
\newblock {\em Phys. Rev. B}, 79(3):033204, 2009.\\

\noindent 
[S6] {Ph.} Hofmann, Ch. S{\o}ndergaard, S.~Agergaard, S.~V. Hoffmann, J.~E. Gayone,
  G.~Zampieri, S.~Lizzit, and A.~Baraldi.
\newblock Unexpected surface sensitivity at high energies in angle-resolved
  photoemission.
\newblock {\em Phys. Rev. B}, 66:245422, 2002.\\

\end{document}